\documentclass[journal]{IEEEtran}
\usepackage{cite}
\usepackage{amsmath,amssymb,amsfonts}
\usepackage{algorithmic}
\usepackage{graphicx}
\usepackage{textcomp}
\usepackage{color}
\usepackage{xcolor}
\usepackage{subfiles}
\usepackage{float}
\usepackage{blindtext}
\usepackage{lipsum}
\usepackage{gensymb}

\usepackage{acronym}
\acrodef{AF}{Atrial Fibrillation}
\acrodef{SR}{Sinus Rhythm}
\acrodef{PVI}{Pulmonary Vein Isolation}
\acrodef{FIRM}{Focal Impulse and Rotor Modulation}
\acrodef{ECGI}{Electrocardiographic Imaging}
\acrodef{LAT}{Local Activation Time}
\acrodef{LA}{Left Atrium}
\acrodef{MRI}{Magnetic Resonance Imaging}
\acrodef{DT-MRI}{Diffusion Tensor Magnetic Resonance Imaging}
\acrodef{DE-MRI}{Delayed-Enhanced Magnetic Resonance Imaging}
\acrodef{LGE-MRI}{Late Gadolinium Enhancement Magnetic Resonance Imaging}
\acrodef{VT}{Ventricular Tachycardia}
\acrodef{CT}{Computed Tomography}
\acrodef{CV}{Conduction Velocity}
\acrodef{ICP}{Iterative Closest Point}
\acrodef{LSPV}{Left Superior Pulmonary Vein}
\acrodef{RSPV}{Right Superior Pulmonary Vein}
\acrodef{LIPV}{Left Inferior Pulmonary Vein}
\acrodef{RIPV}{Right Inferior Pulmonary Vein}
\acrodef{MV}{Mitral Valve}
\acrodef{BB}{Bachmann's Bundle}
\acrodef{LAA}{Left Atrial Appendage}
\acrodef{PS}{Phase Singularity}

\graphicspath{{figure/}}
\begin{document}

\title{Individualization of atrial tachycardia models for clinical applications: Performance of fiber-independent model}

\author{Jiyue He, Arkady Pertsov, John Bullinga, Rahul Mangharam
\thanks{Jiyue He (corresponding author) is with the Department of Electrical and Systems Engineering, University of Pennsylvania, Philadelphia, 19104 USA (e-mail: jiyuehe@seas.upenn.edu).}
\thanks{Arkady Pertsov is with the Department of Pharmacology, Upstate Medical University, Syracuse, USA.}
\thanks{John Bullinga is with Penn Presbyterian Medical Center, Philadelphia, USA.}
\thanks{Rahul Mangharam is with the Department of Electrical and Systems Engineering, University of Pennsylvania, Philadelphia, USA.}
\thanks{Copyright (c) 2023 IEEE. Personal use of this material is permitted. However, permission to use this material for any other purposes must be obtained from the IEEE by sending an email to pubs-permissions@ieee.org.}
}

\maketitle

\begin{abstract}
One of the challenges in the development of patient-specific models of cardiac arrhythmias for clinical applications has been accounting for myocardial fiber organization. The fiber varies significantly from heart to heart, but cannot be directly measured in live tissue. The goal of this paper is to evaluate in-silico the accuracy of left atrium activation maps produced by a fiber-independent (isotropic) model with tuned diffusion coefficients, compares to a model incorporating myocardial fibers with the same geometry. For this study we utilize publicly available DT-MRI data from 7 ex-vivo hearts. The comparison is carried out in 51 cases of focal and rotor arrhythmias located in different regions of the atria. On average, the local activation time accuracy is 96\% for focal and 93\% for rotor arrhythmias. Given its reasonably good performance and the availability of readily accessible data for model tuning in cardiac ablation procedures, the fiber-independent model could be a promising tool for clinical applications.
\end{abstract}


\section{Introduction}
Realistic models of cardiac propagation usually utilize reaction-diffusion equations with highly anisotropic diffusivity tensor which is determined by myocardial fiber organization \cite{Ho2009,Fastl2018}. The ratio of diffusivities along and across fibers could be from 4:1 to as high as 9:1 \cite{Valderrabano2007,franzone1993spread, fenton1998vortex}. The current
best method for acquiring fiber data is via \ac{DT-MRI}, which takes approximately 50 hours to scan \cite{Pashakhanloo2016} and is therefore not clinically practical for cardiac ablation procedures\cite{Parameswaran2021,Hong2020,Calvert2022}. Alternatively, there are rule-based methods for generating synthetic fibers \cite{Roney2019,Grandits2021}. One of the potential approaches for informing the models could be fiber information derived from ex-vivo hearts, however, the individual differences from heart to heart can be very significant and show little spatial correlation. 

The focus of this study is on informing patient-specific models of the left atrium, which harbors severe cardiac arrhythmias such as atrial tachycardia and atrial fibrillation. The analysis of electrical propagation in detailed heart models of the left atrium reveals that the local activation times and gross propagation pattern have relatively low sensitivity to differences in fiber organization \cite{He2022}. This effect is the result of effective isotropization of propagation: Fiber orientations are significantly different across the thickness of the atrial wall, and vary significantly along the atrial surface \cite{He2022}. 

These are two terminologies we use through out the paper: 1) Fiber-inclusive model: A heart model with intrinsic endocardial and epicardial fibers. 2) Fiber-independent model: A heart model without fibers, instead, have tuned diffusion coefficients. Here, we investigate in-silico the performance of fiber-independent model of the left atrium as compared to fiber-inclusive model. 

The major advantage of a fiber-independent model is that it only requires an atrial geometry, which can be readily obtained in clinical setting using either \ac{CT} imaging \cite{kak2001} or electroanatomical mapping \cite{Bhakta2008,He2019}. Such isotropic heart models without fibers have been used previously \cite{Virag2002, Blanc2001, Dam2003}. However, until now there has been no rigorous evaluation of the accuracy. 

There are three main contributions in this paper. First, we conduct comprehensive evaluations of the fiber-independent model for atrial tachycardia, demonstrating its ability to produce highly accurate activation patterns in focal and rotor arrhythmias. Second, we derive a method for tuning diffusion coefficients of the fiber-independent model, enabling it to compensate for the absence of fibers and become patient-specific. Third, we provide detailed explanations of how fibers affect the activation patterns and phase singularity trajectory, clarifying why the fiber-independent model performs well and identifying the circumstances under which it may not perform optimally.

In the following sections, we first describe how to construct a fiber-independent heart model. Then we show comparisons of focal and rotor arrhythmias between fiber-independent and fiber-inclusive models. In the discussion, we explain the effects of fibers and why the fiber-independent model performs well. Lastly, we list the limitations of our work and summarize our findings.

\section{Method}
\subsection{Overview}
\begin{figure*}[!ht]
\centering
\includegraphics[width = 0.95\textwidth]{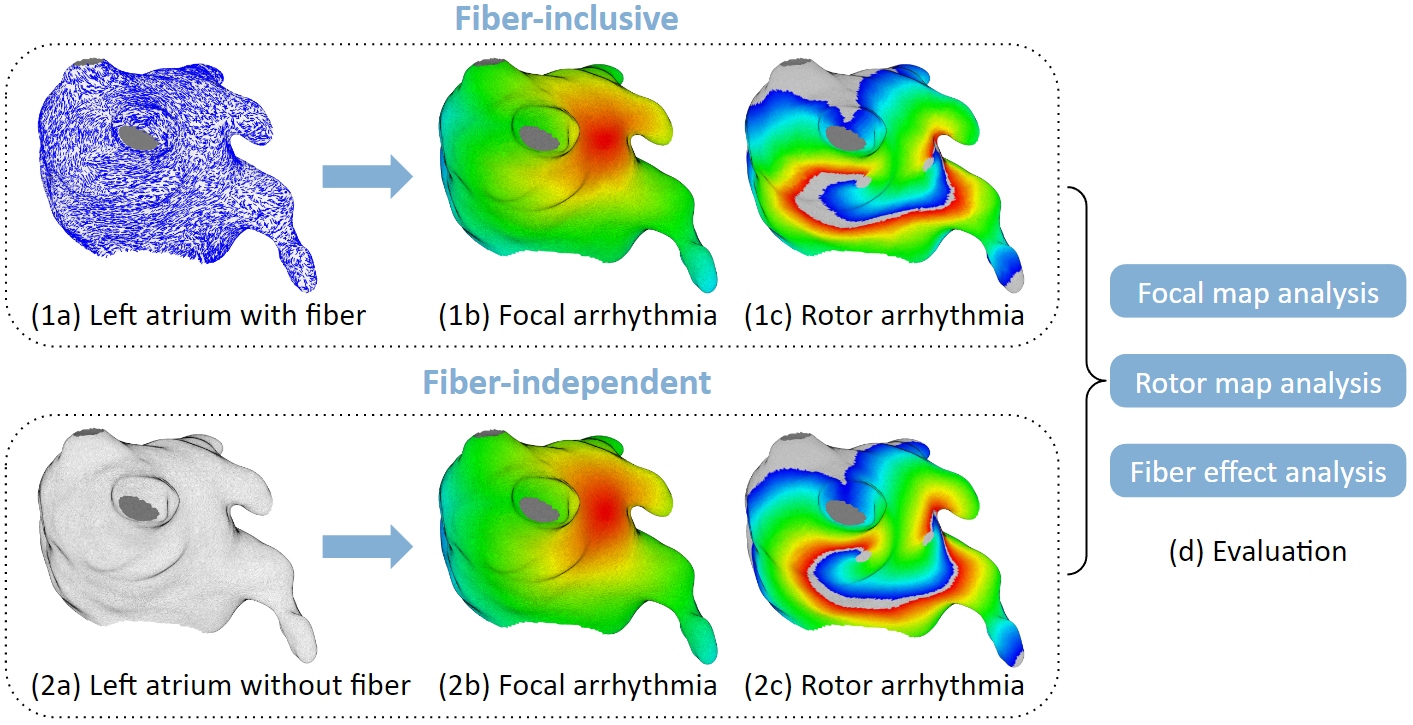}
\caption{Overview. (1a,1b,1c) In-silico experiments on the left atria with a fiber-inclusive model. This model includes endocardial and epicardial fibers, but for artistic reason, only one fiber layer is shown in (1a). (2a,2b,2c) In-silico experiments with a fiber-independent model. This model does not have fiber data, instead, it has tuned diffusion coefficients. (d) Evaluate the fiber-independent model by comparing activation patterns of the focal and rotor arrhythmias to the fiber-inclusive model.}
\label{fig:overview}
\end{figure*}

We utilize a \ac{DT-MRI} database of human atria containing high-resolution information about atrial geometry and fiber organization \cite{Roney2021}. There are 7 left atria (LA$_1$, LA$_2$, …, LA$_7$) each consists of an endocardium 3D triangular mesh (M$_{endo1}$, M$_{endo2}$, …, M$_{endo7}$) and an epicardium mesh (M$_{epi1}$, M$_{epi2}$, …, M$_{epi7}$). The fiber data (F$_1$, F$_2$…, F$_7$) consists of endocardium fibers (F$_{endo1}$, F$_{endo2}$…, F$_{endo7}$) and epicardium fibers (F$_{epi1}$, F$_{epi2}$, …, F$_{epi7}$).

Using these data we create a set of 7 individualized 3D models of the left atrium with intrinsic fiber organization (fiber-inclusive models), and their respective isotropic analogs with the same anatomy and tuned diffusion coefficients (fiber-independent models). To thoroughly compare the performance, we initiate identical focal and rotor arrhythmias in both models, and compare the accuracy of the activation patterns. The comparison is carried out for all 7 left atria: 6 different locations for focal arrhythmias and 1 rotor arrhythmia for each left atrium, and 2 additional rotor arrhythmias for one of the left atria, for a total of 51 cases. The overall approach is outlined in Fig. \ref{fig:overview}.

\subsection{Heart model equations}
To simulate the action potential propagation we employ the mono-domain Mitchell-Schaeffer equations as described in \cite{Mitchell2003,Corrado2018_2}. Computational efficiency of this model makes it particularly useful for large scale 3D numerical simulations, which would not be practical with the use of detailed ionic models \cite{DiFrancesco1985,Shaw1997}.

\begin{align}
\label{eq:mitchell schaeffer}
\begin{split}
\frac{du}{dt}&=\frac{hu^2(1-u)}{\tau _{in}}-\frac{u}{\tau _{out}}+J+\bigtriangledown \cdot (D\bigtriangledown u)
\\
\frac{dh}{dt}&=\left\{\begin{matrix}\ \frac{1-h}{\tau_{open}} \ \text{if}\ u<u_{gate} \\ \ \frac{-h}{\tau_{close}} \ \text{if}\ u \geq u_{gate}\end{matrix}\right.
\end{split}
\end{align}

The variables are as follows: $u$ is the transmembrane voltage and $h$ is an inactivation gating variable for the inward current. $\tau_{in}, \tau_{close}, \tau_{out}, \tau_{open}$ and $u_{gate}$ are parameters that control the action potential shape. $J$ is an external current applied locally as impulses to initiate action potential. We specify this impulse to have 10 ms duration and a magnitude of 20. The part $\bigtriangledown \cdot (D\bigtriangledown u)$ is the diffusion term, responsible for action potential propagation.

For fiber-inclusive models, fiber is introduced via a $3 \times 3$ diffusion tensor $D$ according to \eqref{eq:D},

\begin{align}
\label{eq:D}
D = d \left ( rI+\left ( 1-r \right )ff^\top \right )
\end{align}

Here $d$ is the diffusion coefficient. $r$ is the anisotropy ratio, a ratio of fiber's transverse to longitudinal diffusion coefficients, or the ratio of transverse to longitudinal conduction velocities squared ($r = d_{T}/d_{L}= (CV_{T}/CV_{L})^2$). $I$ is a $3 \times 3$ identity matrix, and $f$ is a $3 \times 1$ unit vector pointing along the fiber direction \cite{Elaff2018}. 

The parameters values are shown in Table \ref{tb:parameter} \cite{Cabrera2017, Roney2019}. With these parameter values, the \ac{CV} is about 0.7 m/s, which is a typical value for the atrium \cite{Harrild2000}. The action potential duration is about 160 ms, which is among the physical observations \cite{Pandit2018}. The rotor arrhythmia has action potential cycle length of about 200 ms, a typical value observed in clinical atrial flutter ablations.

\begin{table}[!ht]
    \centering
    \caption{Parameter values}
    \begin{tabular}{ ccccccc }
    \hline
    $\tau_{in}$ & $\tau_{out}$ & $\tau_{open}$ & $\tau_{close}$ & $u_{gate}$ \\\hline
    0.3 ms & 6 ms & 120 ms & 80 ms & 0.13 mV \\ 
    \hline
    \end{tabular}
    \label{tb:parameter}
    \begin{flushleft}
    Parameter values for focal and rotor arrhythmia simulations. For fiber-inclusive models, we set $r$ = 0.2 and $d$=1. For fiber-independent models, we set $r$ = 1 (which removes fibers), and then tune the value of $d$.
    \end{flushleft}
\end{table}

The average surface area of a left atrium is 33,000 mm$^2$, number of mesh vertices is 86,000, number of mesh faces is 171,000, the triangle edge length is 0.67 mm, and the bounding box size of the 7 atria is 106 mm $\times$ 118 mm $\times$ 116 mm. We create Cartesian nodes (with 1 mm spacing) wrapping around the endocardium mesh. The average number of Cartesian nodes for a left atrium is 94,000.

The heart model equation \eqref{eq:mitchell schaeffer} is solved on the Cartesian nodes. The nodes' spatial resolution and the solver's temporal resolution is chosen to be adequate for accurate simulation without being computationally demanding, details please refer to Appendix \ref{app:resolution}. Then the values of the nodes are projected to mesh vertices.

For fiber-inclusive models, we register both endocardial and epicardial fibers to the nodes. Please refer to our previous paper for details \cite{He2022}.

For fiber-independent models, $r$ is set to 1, therefore, fibers are removed. Equation \eqref{eq:D} reduces to $D=dI$, which is $d$ times a $3\times3$ identity matrix. Here the value of $d$ will be tuned to patient-specific value, details on the implementation see Section \ref{sec:tune diffusion}.

The differential equations \eqref{eq:mitchell schaeffer} are solved using the explicit Euler method on the Cartesian nodes. We follow \cite{McFarlane2010} that assumed no-flux boundary conditions and use a 19-node stencil. For more details, please refer to Appendix \ref{app:solve heart model}. The model is implemented in MATLAB (MathWorks, Natick, Massachusetts, United States) and accelerated with GPU computing using CUDA kernels (Nvidia, Santa Clara, California, United States).

\subsection{Arrhythmia simulations setup}
\begin{figure}[!ht]
\centering
\includegraphics[width = 0.45\textwidth]{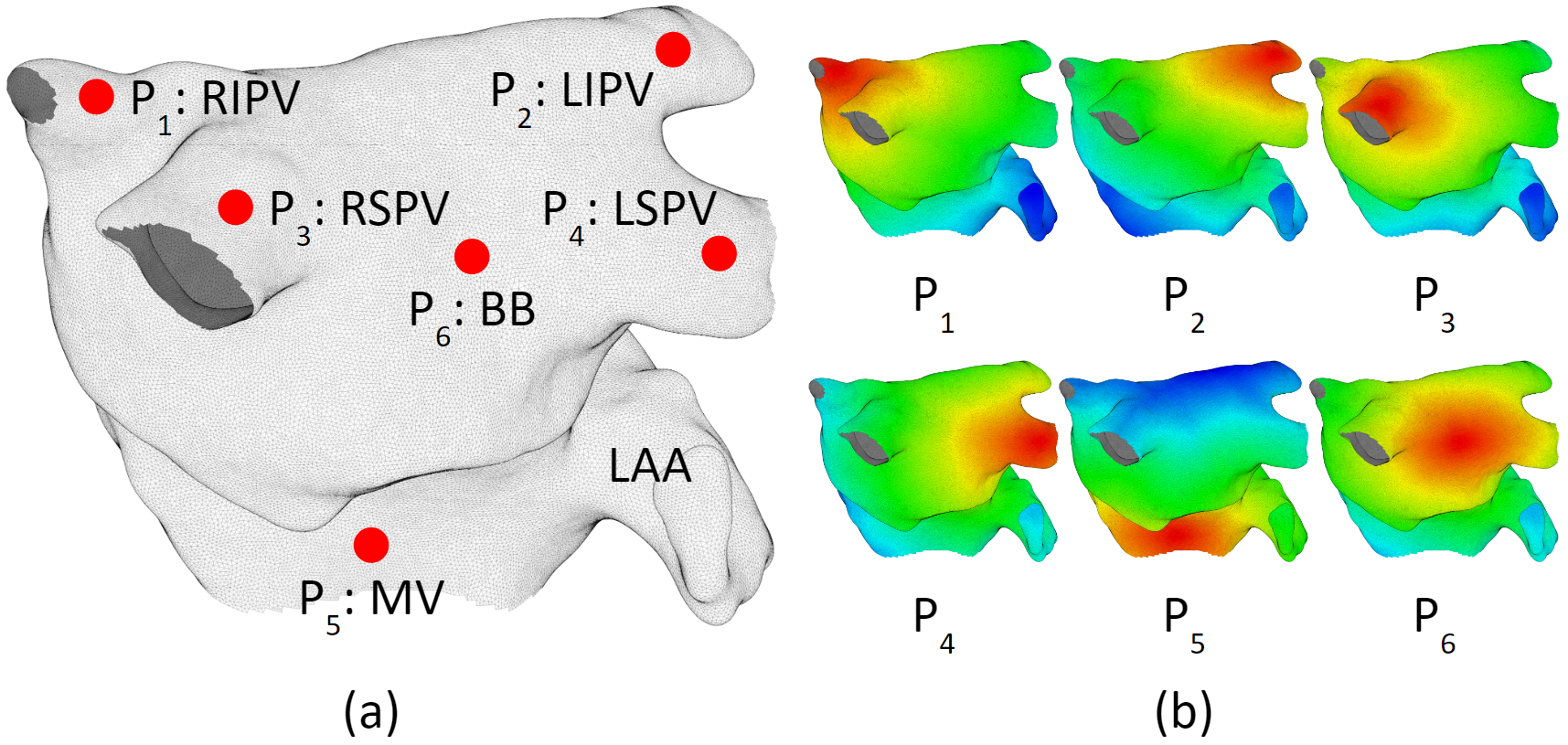}
\caption{Focal arrhythmia simulations setup. (LA$_3$ is shown here.) (a) Pacing sites for focal arrhythmia simulation. (b) Examples of activation maps obtained for each of the pacing locations. Red represents early activation and blue represents late activation. The focal source is located at the center of the red region. LAA: left atrial appendage.}
\label{fig:focal setup}
\end{figure}

We specify 6 different focal arrhythmia locations on each atrium. The pacing sites are shown in Fig. \ref{fig:focal setup}. P$_1$: \ac{RIPV}, P$_2$: \ac{LIPV}, P$_3$: \ac{RSPV}, P$_4$: \ac{LSPV}, P$_5$: \ac{MV}, and P$_6$: \ac{BB}. These locations are chosen because they are clinically identifiable, also they cover a wide variety of scenarios.

\begin{figure}[!ht]
\centering
\includegraphics[width = 0.45\textwidth]{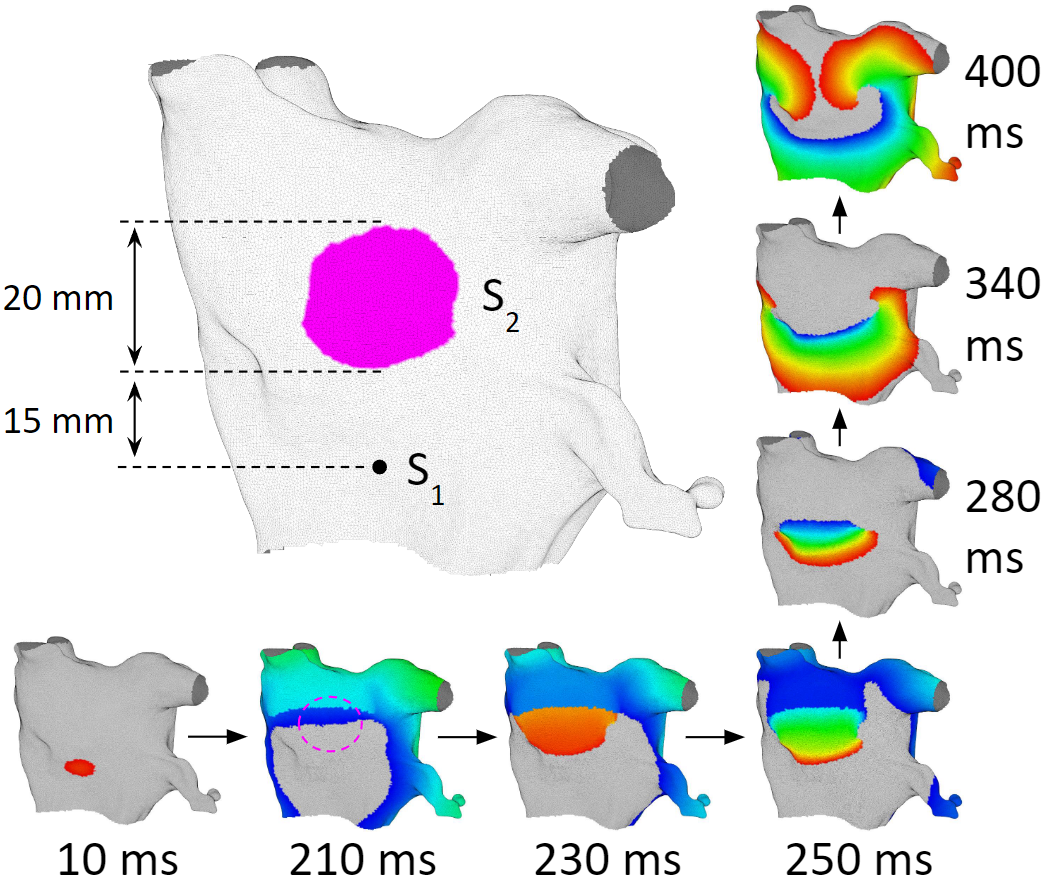}
\caption{Rotor arrhythmia simulations setup. (LA$_1$ is shown here.) S$_1$ and S$_2$ are the locations of stimuli. S$_1$ stimulus is applied focally at the \ac{MV} region, initiating a propagating wave (see the 10 ms snapshot). S$_2$ stimulus depolarizes a larger area (the magenta  region). It is applied after S$_1$ at the tail of the propagating action potential. 210 and 230 ms snapshots show the activation right before and after the S$_2$ stimulus, respectively. Magenta dashed circle in the 210 ms snapshot indicates the S$_2$ area. Snapshots 250-400 ms show the formation of a pair of rotors at the intersection of S$_2$ with the tail of S$_1$ wave.}
\label{fig:rotor setup}
\end{figure}

For rotor arrhythmia simulations, we use a method illustrated in Fig. \ref{fig:rotor setup}, which produces a pair of rotors in a desired location. The method utilizes two electrical stimuli S$_1$ and S$_2$: S$_1$ stimulus is applied focally at the \ac{MV} region, initiating a propagating wave, S$_2$ stimulus depolarizes a larger area (magenta) and is applied after S$_1$ at the tail of the propagating action potential. The size and the position of S$_2$ defines the rotor location. To make the rotor stable in some of the simulations, we place small (3-5 mm) non-conducting fibrotic patches in the rotation centers as anchors \cite{Zemlin2012}.

\subsection{Fiber-independent modeling}
\label{sec:tune diffusion}

\begin{figure*}[!ht]
\centering
\includegraphics[width = 0.95\textwidth]{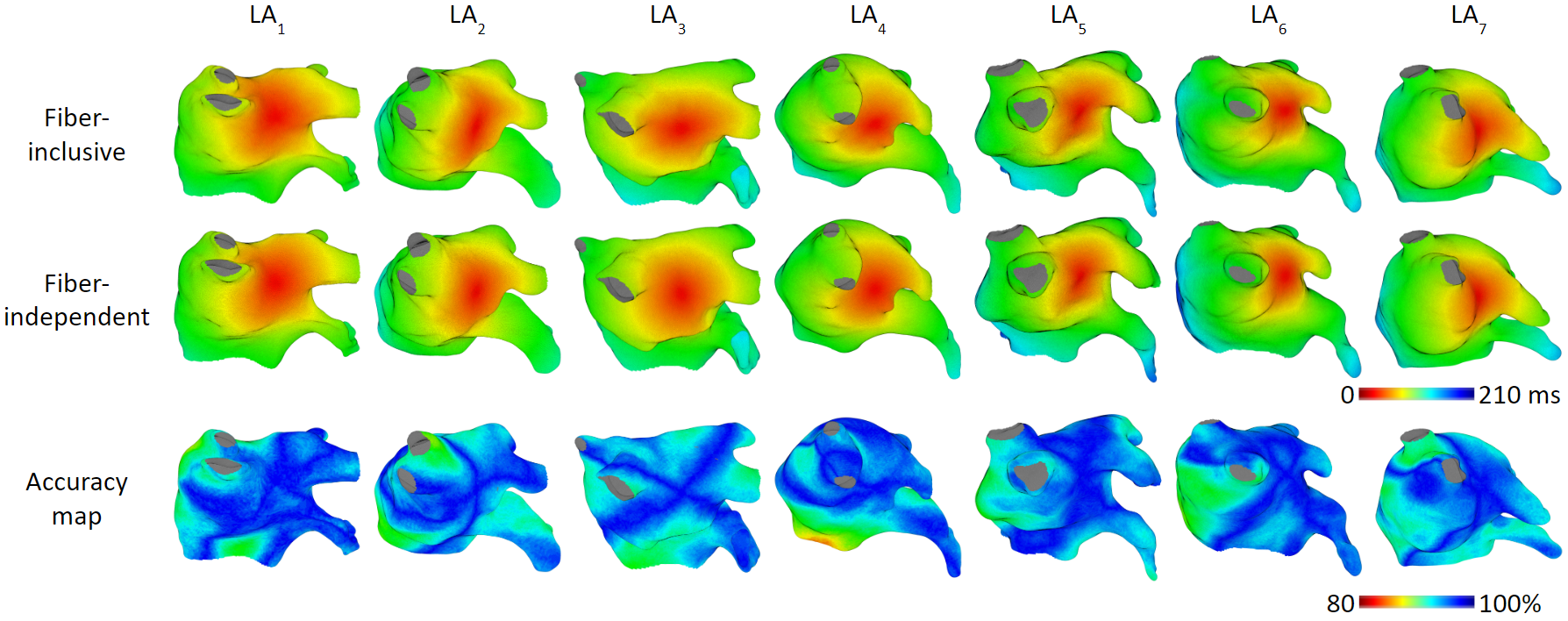}
\caption{Focal arrhythmia comparison: A total of 42 focal arrhythmias are generated (7 left atria each with 6 focal pacing scenarios), here shows results of focal pacing at location P$_6$. Row 1 and 2 show local activation time maps of fiber-inclusive and fiber-independent models respectively. We can see that the red regions in row 2 are more rounded, because the model assumes isotropic conduction; while red regions in row 1 are skewed by fiber orientations. Row 3 shows the accuracy maps calculated according to Equation \eqref{eq:accuracy}, with 100\% meaning no difference between the two models.}
\label{fig:focal lat map}
\end{figure*}

The fiber-independent model uses an isotropic spatially uniform diffusion. The diffusion coefficient is tuned to approximate the effect of fiber organization on tissue conductivity. For tuning we used the P$_5$ activation map, generated using the respective fiber-inclusive model as the ground truth. The selection of the P$_5$, as opposed to other pacing locations, is intended to simulate the tuning procedure in clinical setting. P$_5$ is a point in the \ac{MV} region, where the coronary sinus catheter resides during ablation procedure. An electroanatomical map \cite{He2019} with pacing from the coronary sinus can be acquired at the beginning of the ablation procedure using Pentaray or Lasso catheters. With systems such as the Carto3 System (Biosense Webster, Inc. Irvine, California, United States), it typically takes a physician 3 minutes to measure about 2,500 sample points that cover the entire left atrium.





Tuning of the diffusion coefficient in fiber-independent model is done in three steps. First, we generate a surrogate clinical activation map using the ground truth fiber-inclusive model for a given atrium, which we pace from P$_5$ as described above.  Then we run a fiber-independent model of the same atrium and generate the first iteration of P$_5$ activation map with some initial guess value of the diffusion coefficient $d_0$. The resulting two maps are used to tune the diffusion coefficient and generate the second (final) iteration of the fiber-independent model. To calculate the tuned value of the diffusion coefficient we use the following formula:
\begin{align}
\label{eq:d_tuned_all_points}
d_{tuned} = d_{0} \frac{1}{N} \sum_{n=1}^{N} \left(\frac{T_{0,n}}{T_{1,n}}\right)^2 
\end{align}
Where $d_0$ is the initial guess value of the diffusion coefficient, $N$ is the total number of nodes. $T_{0,n}$ and $T_{1,n}$ are the traveling times for the activation to get to this $n$-th node from the pacing site (P$_5$) in the fiber-independent and fiber-inclusive models, respectively. By using such tuning we minimize the differences between local activation times in fiber-independent and fiber-inclusive models.

The formula \eqref{eq:d_tuned_all_points} is based on the notion that in any location the activation time is inversely proportional to conduction velocity ($ T\propto1/CV $) and that in the reaction diffusion systems the ratio of conduction velocities ${CV_0}$ and ${CV_1}$ evaluated at different diffusion coefficients $d_0$ and $d_1$ is equal to square root of the ratio of diffusion coefficients $CV_0/CV_1$ = $(d_0/d_1)^{1/2}$. Accordingly, the ratio of conduction times at a given location at different diffusion coefficients can be expressed as $T_0/T_1$ = $CV_1/CV_0$ = $(d_1/d_0)^{1/2}$. After a simple transformation one can obtain a formula linking the values of diffusion coefficients to activation times $d_1/d_0$ = $(T_0/T_1)^2$ that we utilize for tuning the fiber-independent model.

For each atrium, the tuning is done only once, and the tuned fiber-independent model is used for all arrhythmia scenarios mentioned above.

\subsection{Evaluating fiber-independent model performance}
We calculate the absolute \ac{LAT} error and accuracy of every vertices between the two models. The error is defined according to \eqref{eq:error},
\begin{align}
\label{eq:error}
\text{Error} = \frac{1}{M} \sum_{m=1}^{M} \left| LAT_{1,m} - LAT_{2,m} \right|
\end{align}
where $M$ is the total number of vertices, and $LAT_{1,m}$ and $LAT_{2,m}$ are the \ac{LAT} values of the $m$-th vertex from fiber-inclusive and fiber-independent model respectively.

Accuracy at a vertex is defined according to \eqref{eq:accuracy},
\begin{align}
\label{eq:accuracy}
\text{Accuracy} = \left(1 - \frac{\left| LAT_{1} - LAT_{2} \right|}{T} \right) \times 100\%
\end{align}
where $LAT_{1}$ and $LAT_{2}$ are the \ac{LAT} values of a same vertex of the fiber-inclusive and fiber-independent model respectively. $T$ is calculated from fiber-inclusive model simulation data. For focal arrhythmia, $T=Range(LAT)$, it represents the time for the activation wave to travel across the entire left atrium. For rotor arrhythmia, $T=rotor\ cycle\ length$, it represents the time for the rotor to rotate 360 degrees.

\section{Results}
\subsection{Fiber-independent model accuracy on focal arrhythmias}
We generate 42 focal arrhythmias: 7 left atria each with 6 pacing scenarios. Fig. \ref{fig:focal lat map} shows some of the results. Observe the red regions of row 1 (fiber-inclusive models): In LA$_2$ the red region stretched vertically and in LA$_5$ it stretched in the 1 and 7 o'clock direction. Upon further investigation, we find that these directions of stretch are parallel with fiber orientations. On the other hand, the red regions in row 2 (fiber-independent models) are more rounded, mainly because these models assume isotropic conduction. 

\subsubsection{Local activation time error}
\begin{figure}[!ht]
\centering
\includegraphics[width = 0.45\textwidth]{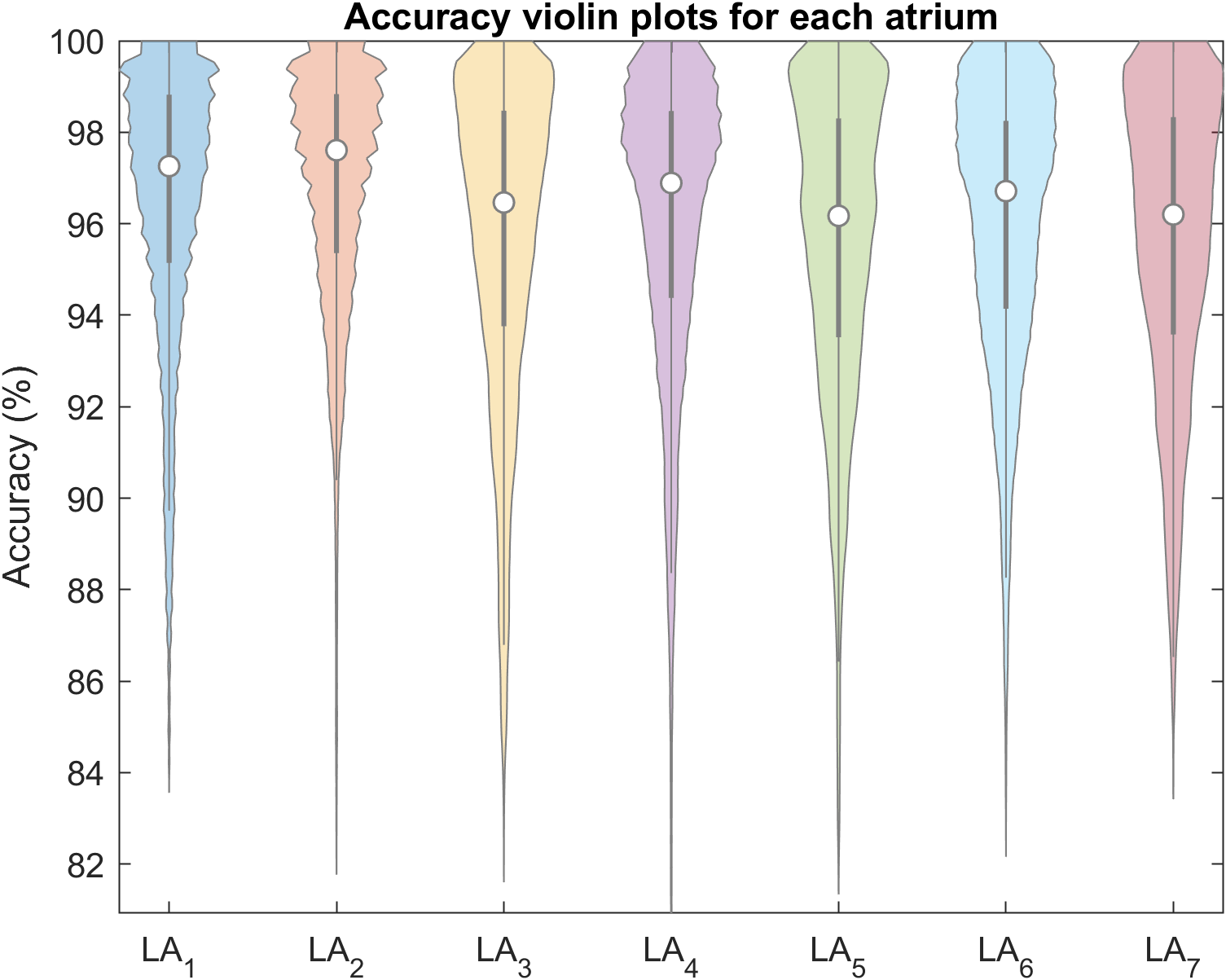}
\caption{Violin plots of focal arrhythmia local activation time accuracy of each left atrium. The overall average accuracy is 96\%.}
\label{fig:focal accuracy}
\end{figure}

Results are summarized in Table \ref{tb:lat err focal arrhythmia}. The overall error is 9$\pm$7 ms, or 96\% accuracy. Details of the accuracy is shown in Fig. \ref{fig:focal accuracy}.  

\begin{table}[!ht]
    \scriptsize
    \centering
    \caption{Local activation time error (ms)}
    \begin{tabular}{ c | c c c c c c c | c }
    \hline
 & LA$_1$ & LA$_2$ & LA$_3$ & LA$_4$ & LA$_5$ & LA$_6$ & LA$_7$ & \textbf{Avg} \\ \hline
P$_1$ & 4$\pm$3 & 6$\pm$6 & 6$\pm$5 & 11$\pm$7 & 5$\pm$4 & 10$\pm$8 & 14$\pm$10 & \textbf{8$\pm$7} \\ 
P$_2$ & 6$\pm$4 & 7$\pm$6 & 14$\pm$9 & 9$\pm$8 & 14$\pm$8 & 9$\pm$6 & 9$\pm$6 & \textbf{10$\pm$8} \\ 
P$_3$ & 8$\pm$6 & 4$\pm$3 & 7$\pm$5 & 9$\pm$8 & 13$\pm$7 & 10$\pm$7 & 9$\pm$9 & \textbf{9$\pm$7} \\ 
P$_4$ & 9$\pm$6 & 7$\pm$6 & 9$\pm$7 & 6$\pm$5 & 15$\pm$8 & 11$\pm$9 & 9$\pm$6 & \textbf{10$\pm$7} \\ 
P$_5$ & 5$\pm$3 & 5$\pm$4 & 10$\pm$8 & 5$\pm$4 & 6$\pm$4 & 8$\pm$5 & 7$\pm$5 & \textbf{7$\pm$5} \\ 
P$_6$ & 7$\pm$6 & 7$\pm$5 & 9$\pm$6 & 8$\pm$6 & 8$\pm$6 & 9$\pm$6 & 9$\pm$6 & \textbf{8$\pm$6} \\ \hline 
\textbf{Avg} & \textbf{6$\pm$5} & \textbf{6$\pm$5} & \textbf{9$\pm$7} & \textbf{8$\pm$7} & \textbf{10$\pm$8} & \textbf{9$\pm$7} & \textbf{10$\pm$8} & \textbf{9$\pm$7} \\ \hline 
    \end{tabular}
    \label{tb:lat err focal arrhythmia}
    \begin{flushleft}
    \footnotesize
    Focal arrhythmia LAT errors between fiber-inclusive and fiber-independent models. Average error for all scenarios is 9$\pm$7 ms.
    \end{flushleft}
\end{table}

\subsubsection{Latest activation location error}
One of the clinically relevant characteristics  that we use to evaluate the predictive capabilities of fiber-independent model is the accuracy of prediction of the regions with the latest activation (within 10 ms of the maximum \ac{LAT}.)

\begin{figure}[!ht]
\centering
\includegraphics[width = 0.35\textwidth]{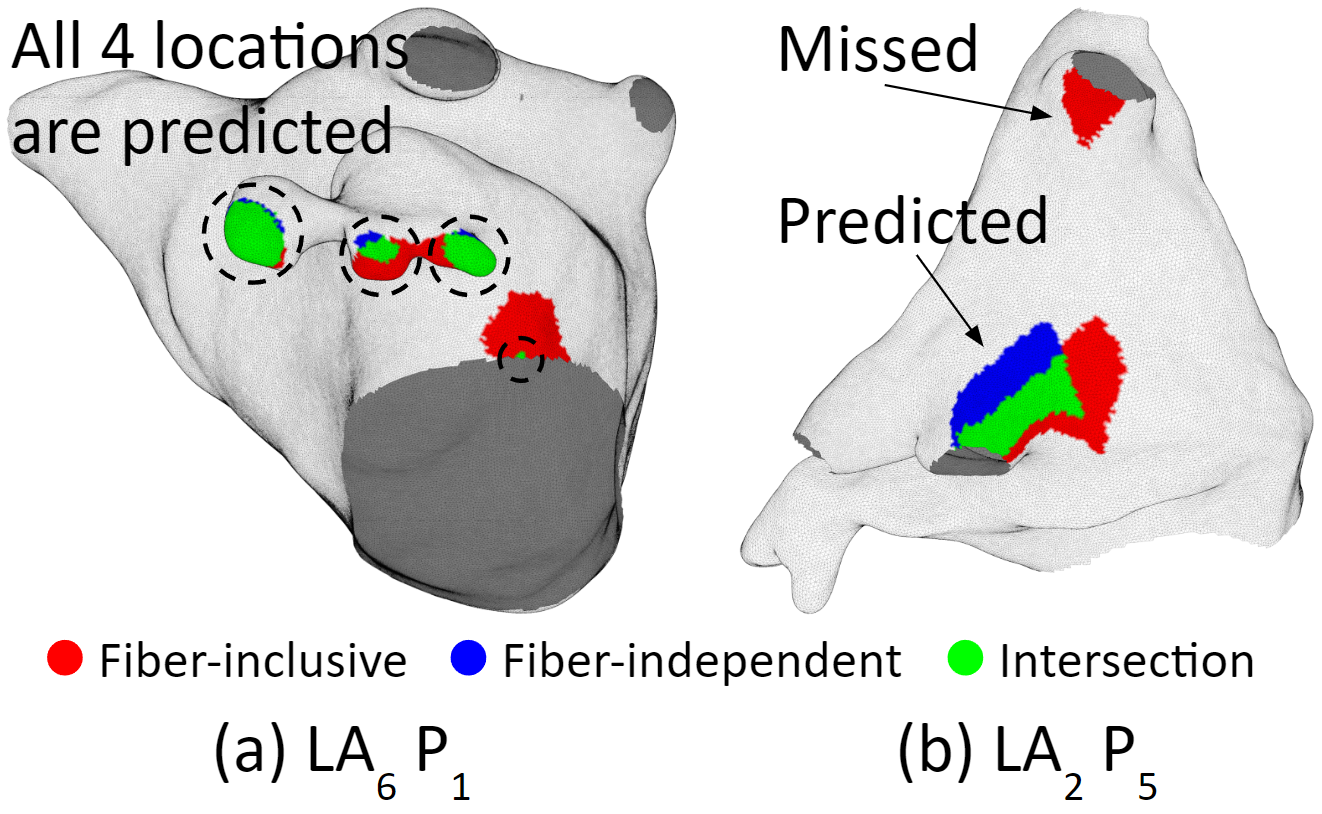}
\caption{Focal arrhythmia latest activation location plots. Red and blue represent results of fiber-inclusive and fiber-independent model respectively, and green is the overlapping area. (a) Latest activation locations of P$_1$ focal arrhythmia on LA$_6$. We can see that the regions of the two models overlap well, resulting a small error. (b) The case of P$_5$ focal arrhythmia on LA$_2$. We can see that the error is small in the red-blue-green overlapping region, however, the fiber-independent model did not predict the upper red region. This happened 4 times, accounted for 9.5\% of all 42 scenarios.}
\label{fig:latest activation location}
\end{figure}

Fig. \ref{fig:latest activation location} shows the location of these regions in LA$_6$ and LA$_2$ in fiber inclusive and fiber-independent models for focal arrhythmias P$_1$ and P$_5$, respectively. The Euclidean distance between fiber-inclusive and fiber-independent model's latest activation locations are summarized in Table \ref{tb:latest activation location}. The average distance is 7.5 mm (the size of the left atrium is 106 mm $\times$118 mm$\times$116 mm, 7.5 is 6.4\% of 118). 

In most cases, the fiber-independent model predicts well: See Fig. \ref{fig:latest activation location}(a), all 4 locations are predicted. However, we observe in some cases that the fiber-independent model fail to predict all of the latest activation locations: See Fig. \ref{fig:latest activation location}(b). This happened 4 times, or 9.5\% of all 42 scenarios. 

\begin{table}[!ht]
    \centering
    \caption{Latest activation location difference (mm)}
    \begin{tabular}{ c | c c c c c c c | c }
    \hline
 & LA$_1$ & LA$_2$ & LA$_3$ & LA$_4$ & LA$_5$ & LA$_6$ & LA$_7$ & \textbf{Avg} \\ \hline
P$_1$ & 0.5 & 5.8 & 0.5 & 4.8 & 7.9 & 4.8 & 1.1 & \textbf{3.6} \\ 
P$_2$ & 7.1 & 12.3 & 4.7 & 5.8 & 7.4 & 3.9 & 5.7 & \textbf{6.7} \\ 
P$_3$ & 6.1 & 7.7 & 3.7 & 3.2 & 3.4 & 5.3 & 16.3 & \textbf{6.5} \\ 
P$_4$ & 2.6 & 8.8 & 3.7 & 31.9 & 5.9 & 5.2 & 14.1 & \textbf{10.3} \\ 
P$_5$ & 18.2 & 8.0 & 21.4 & 33.3 & 9.1 & 10.6 & 1.7 & \textbf{14.6} \\ 
P$_6$ & 2.5 & 2.1 & 3.2 & 7.0 & 0.4 & 7.4 & 1.8 & \textbf{3.5} \\ \hline 
\textbf{Avg} & \textbf{6.2} & \textbf{7.4} & \textbf{6.2} & \textbf{14.3} & \textbf{5.7} & \textbf{6.2} & \textbf{6.8} & \textbf{7.5} \\ \hline 
    \end{tabular}
    \label{tb:latest activation location}
    \begin{flushleft}
    Focal arrhythmia latest activation location difference between fiber-inclusive and fiber-independent models. The overall average is 7.5 mm.
    \end{flushleft}
\end{table}

\subsection{Fiber-independent model accuracy on rotor arrhythmias}
\begin{figure*}[!ht]
\centering
\includegraphics[width = 0.95\textwidth]{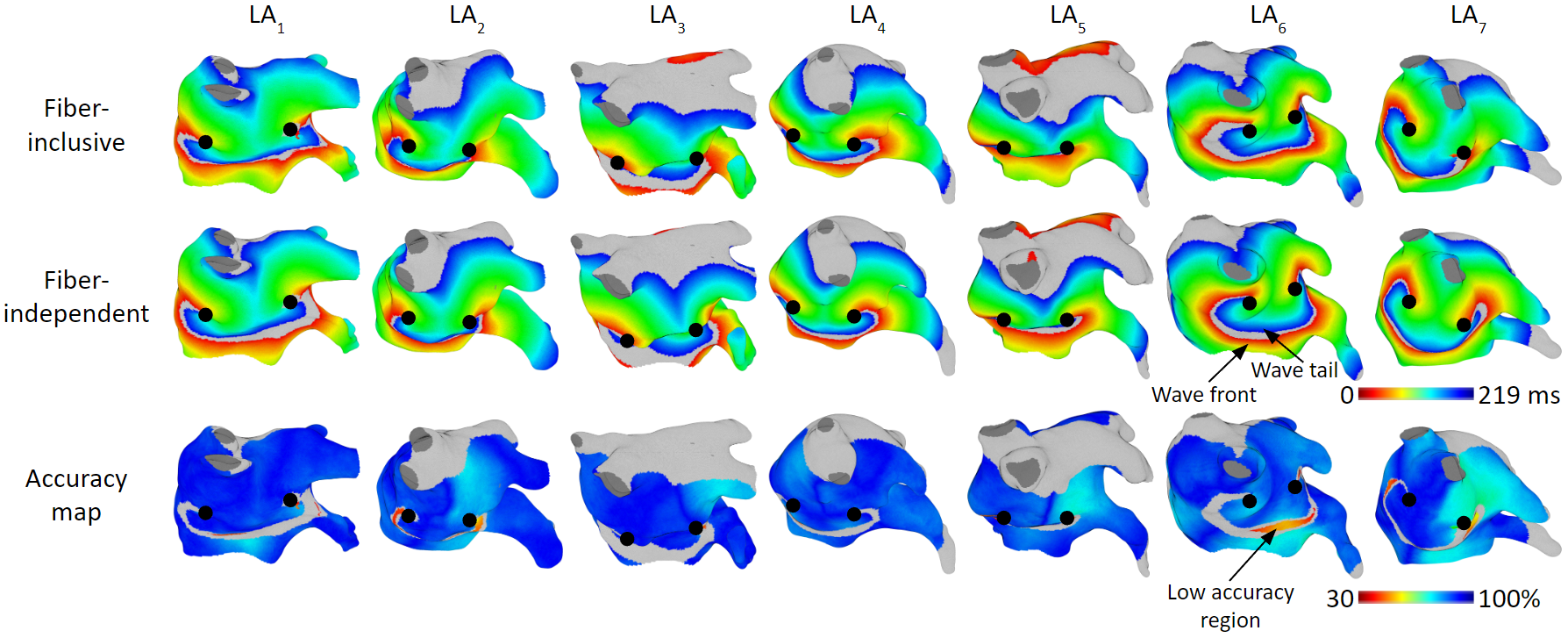}
\caption{Rotor arrhythmia comparison. A pair of rotors are generated on each left atrium. These rotors were made stable by placing small non-conducting patches as anchors (marked with black dots) at the rotation center, which allowed us to compare the activation patterns of the rotor arrhythmias between models. Row 1 and 2 show local activation time maps of fiber-inclusive and fiber-independent models respectively. Row 3 shows the accuracy maps calculated according to Equation \eqref{eq:accuracy}, with 100\% meaning no difference between the two models.}
\label{fig:rotor lat map}
\end{figure*}

Atrial flutter, macro re-entry and atrial fibrillation can have rotating activation waves, which can be represented as rotors. 
We create stable rotors on the 7 left atria. Fig. \ref{fig:rotor lat map} shows the comparisons. We can see the activation wavefronts are slightly different between the two models, however, the over all activation patterns are similar. Quantitative analysis are shown in Table \ref{tb:lat err rotor arrhythmia}. The overall \ac{LAT} error is 14$\pm$16 ms, or 93\% accuracy.

\begin{table}[!ht]
    \scriptsize
    \centering
    \caption{Local activation time error (ms)}
    \begin{tabular}{ c c c c c c c | c }
    \hline
LA$_1$ & LA$_2$ & LA$_3$ & LA$_4$ & LA$_5$ & LA$_6$ & LA$_7$ & \textbf{Avg} \\ \hline
8$\pm$10 & 13$\pm$16 & 14$\pm$14 & 12$\pm$11 & 14$\pm$18 & 15$\pm$15 & 23$\pm$20 & \textbf{14$\pm$16} \\ \hline 
    \end{tabular}
    \label{tb:lat err rotor arrhythmia}
    \begin{flushleft}
    \footnotesize
    Rotor arrhythmia LAT errors between fiber-inclusive and fiber-independent models. Average error for all scenarios is 14$\pm$16 ms.
    \end{flushleft}
\end{table}

\begin{figure}[!ht]
\centering
\includegraphics[width = 0.45\textwidth]{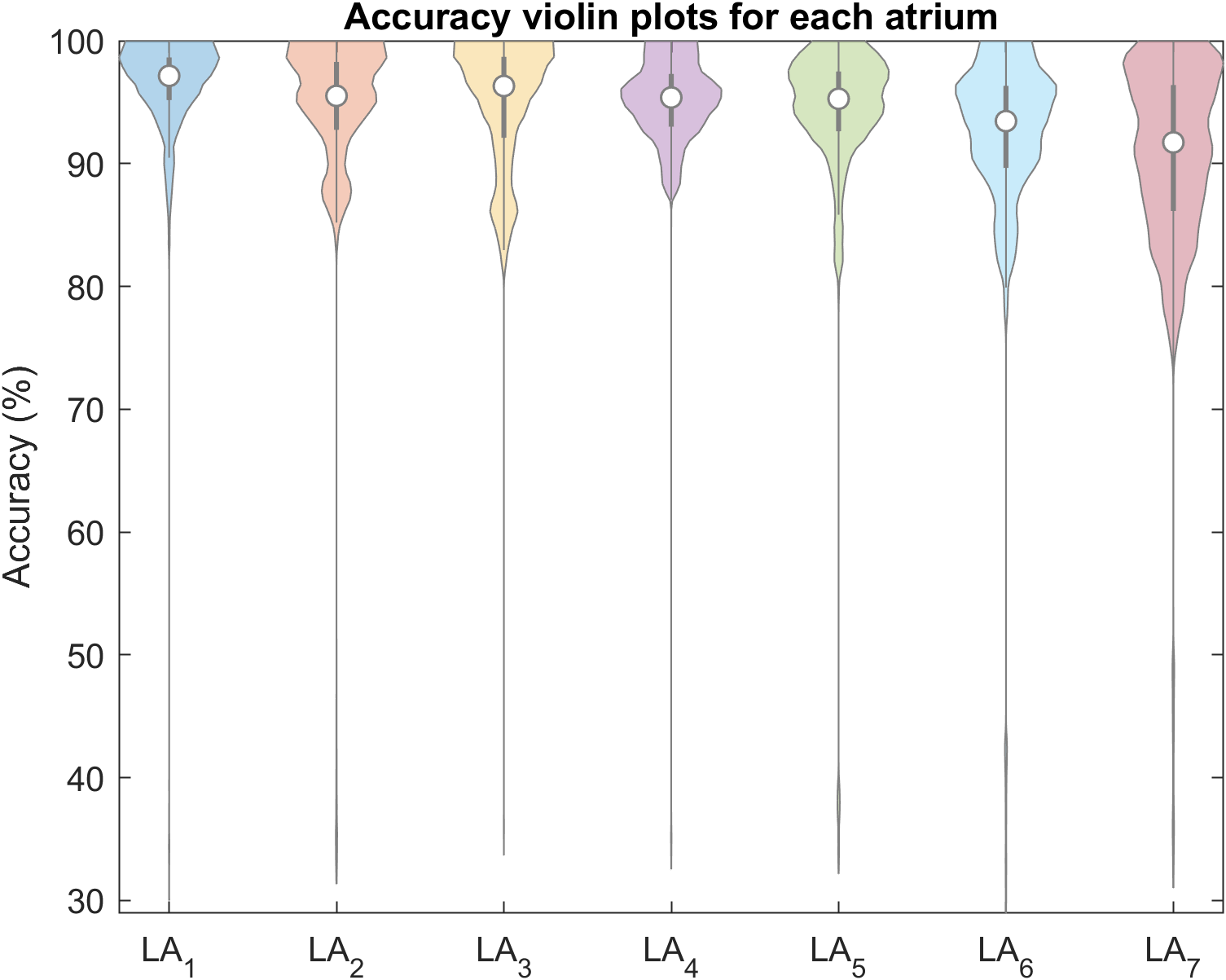}
\caption{Violin plots of rotor arrhythmia local activation time accuracy of each left atrium. The overall average accuracy is 93\%.}
\label{fig:rotor accuracy}
\end{figure}

Details of the accuracy is shown in Fig. \ref{fig:rotor accuracy}. We can see that the majority accuracy are higher than 90\%. However, we notice that there are a small amount of accuracy in the 30\%-40\% range. These low accuracy regions are the red regions in the accuracy maps of Fig. \ref{fig:rotor lat map} (row 3). For example, on LA$_6$, the low accuracy region is marked by an arrow, it is located in between the wave front (red, marked by an arrow) and wave tail (blue, marked by an arrow). Wave front has \ac{LAT} value of 0 ms, wave tail has \ac{LAT} value of 219 ms, these two values are far apart, thus small deviation in the wave front / wave tail location can result in large error, or low accuracy.

\begin{figure}[!ht]
\centering
\includegraphics[width = 0.45\textwidth]{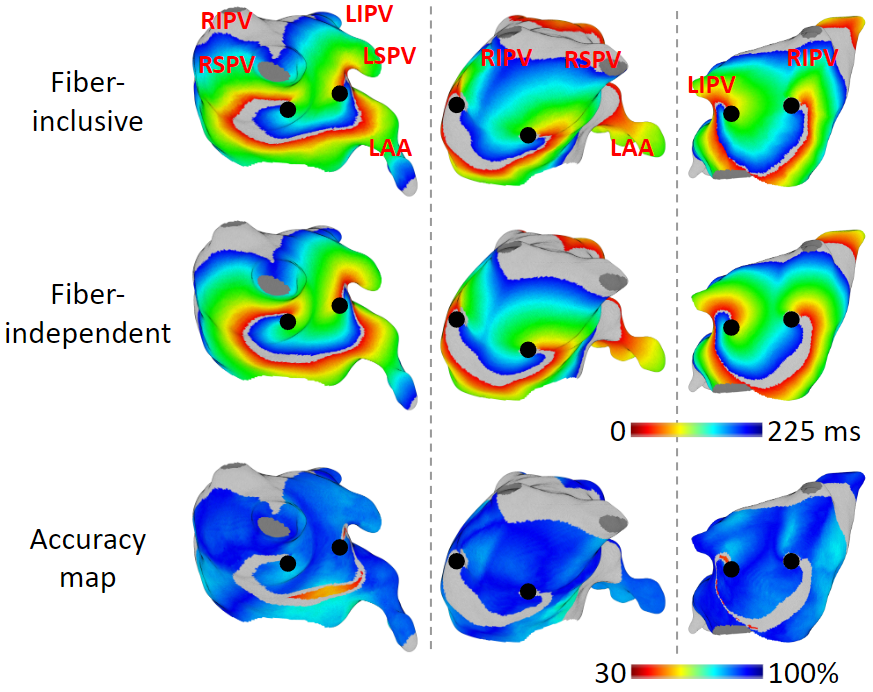}
\caption{Local activation time maps of different rotor arrhythmias on LA$_{6}$. All atria here are the same one but with different view angles. The activation patterns between the two models are similar.}
\label{fig:different rotor location}
\end{figure}

To observe if model performance is different for different rotor locations, we create two more rotor arrhythmias on LA$_{6}$, make a total of three different rotor arrhythmias for this atrium. LA$_{6}$ is chosen, because in Table \ref{tb:lat err rotor arrhythmia}, it has an average performance among the seven atria. As shown in Fig. \ref{fig:different rotor location}, the \ac{LAT} errors are similar: Left column \ac{LAT} error is 15$\pm$15 ms (93\% accuracy), middle column is 15$\pm$12 ms (94\% accuracy), right column is 12$\pm$13 ms (94\% accuracy).

We also find there are slight differences in rotor cycle lengths between the two models. Details are summarized in Table \ref{tb:rotor cycle length}.

\begin{table}[!ht]
    \centering
    \caption{Rotor average cycle length (ms)}
    \begin{tabular}{ c | c c c c c c c | c }
    \hline
         & LA$_1$ & LA$_2$ & LA$_3$ & LA$_4$ & LA$_5$ & LA$_6$ & LA$_7$ & \textbf{Avg} \\ \hline
CL$_1$ & 210 & 223 & 217 & 216 & 233 & 214 & 217 & \textbf{219} \\ 
CL$_2$ & 212 & 215 & 219 & 213 & 227 & 219 & 217 & \textbf{217} \\ \hline 
    \end{tabular}
    \label{tb:rotor cycle length}
    \begin{flushleft}
    Rotor average cycle length. CL$_1$: fiber-inclusive rotor cycle length. CL$_2$: fiber-independent rotor cycle length.
    \end{flushleft}
\end{table}

\section{Discussion}
\subsection{Fiber analysis}
Obtaining accurate and high-resolution fiber data for cardiac ablation procedure is a challenge. One solution is to use synthetic fibers, which can be mathematically generated based on the heart's geometry \cite{Fastl2018, Krueger2011, Labarthe2021, Wachter2015, Saliani2021}. Alternatively, fibers from existing databases can be registered, as it has been found that some patients' fibers can be generalized to many different patients \cite{Roney2021}. However, fibers obtained through these methods do not represent the true fibers. As analyzed below, left atrium fibers vary across patients.

\begin{figure}[!ht]
\centering
\includegraphics[width = 0.45\textwidth]{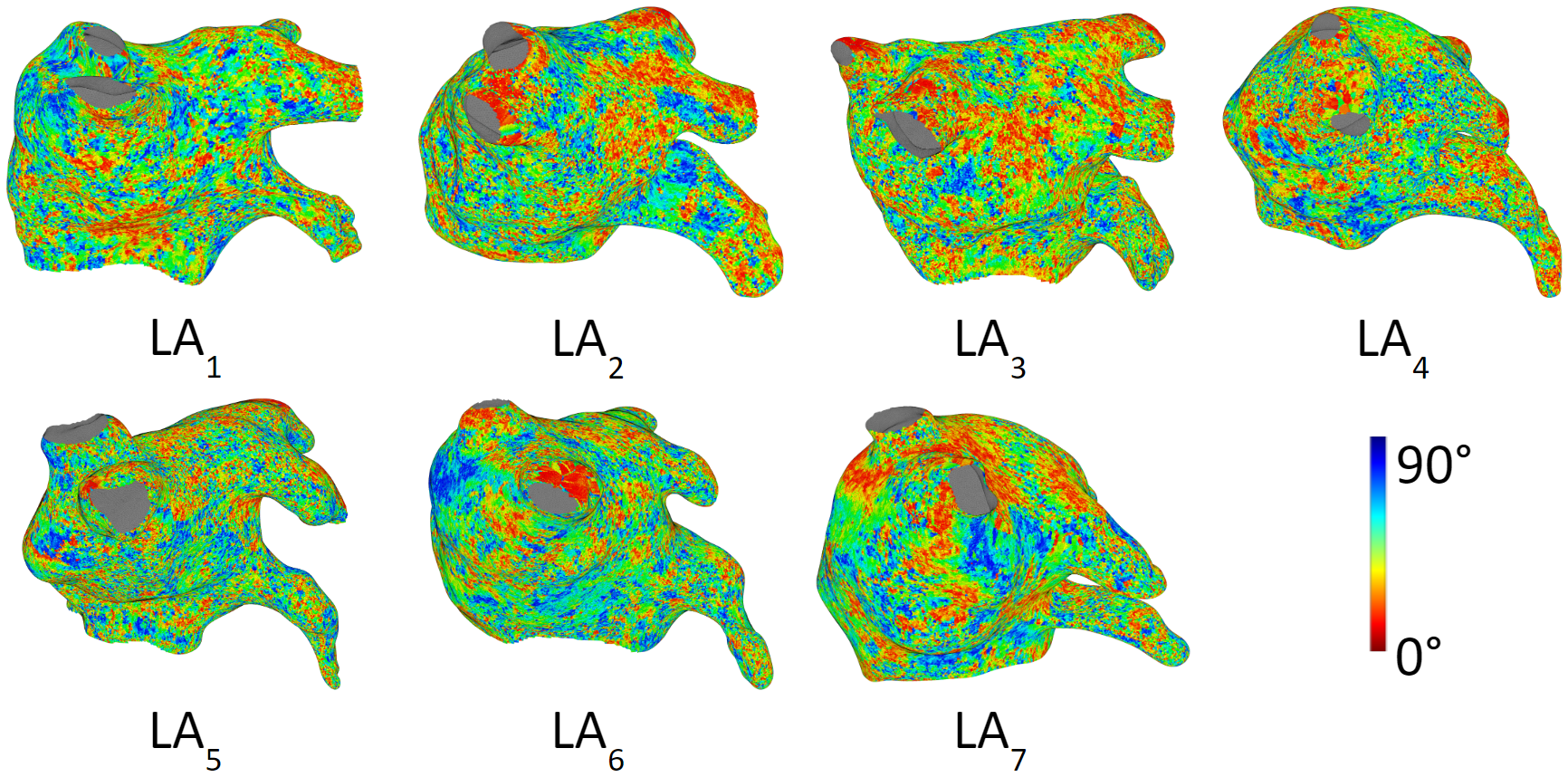}
\caption{Endocardium-epicardium fiber angle difference maps. There is no consistent pattern for the red blue locations within and among different atria.}
\label{fig:fiber angle difference}
\end{figure}

We compute the endocardium-epicardium fiber angle differences $\Delta\theta$ for each left atrium. Fig. \ref{fig:fiber angle difference} shows the $\Delta\theta$ map. Random local color variations indicate lack of large regions with the same $\Delta\theta$ value.

\subsection{Fiber orientation vs activation pattern}
Studies have found that isotropic heart models work relatively well \cite{Virag2002, Ruchat2007, Gray1996}. Our previous research on the effects of fibers on activation patterns found a possible explanation for this phenomenon, which we refer to as the cancellation effect \cite{He2022}.

\begin{figure}[!ht]
\centering
\includegraphics[width = 0.45\textwidth]{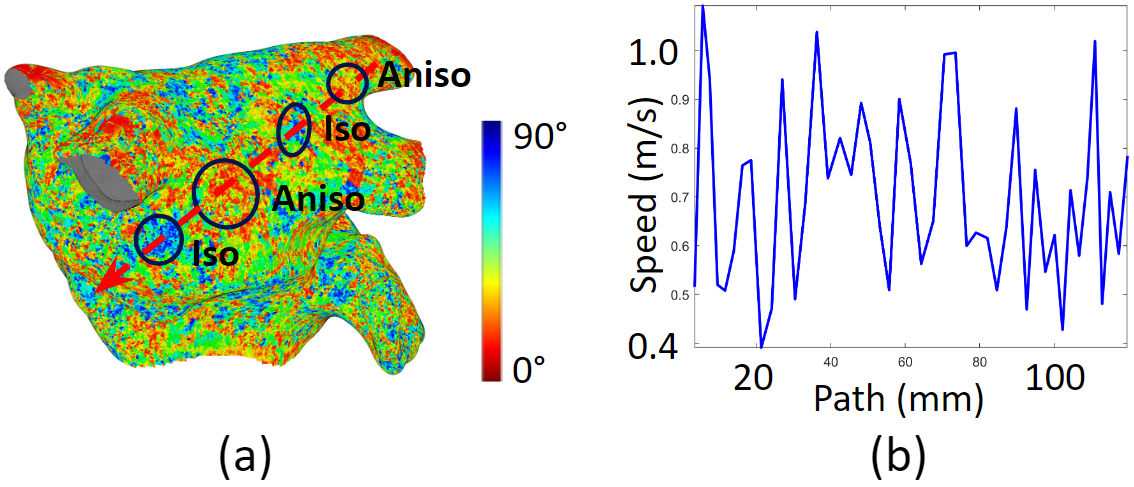}
\caption{Left atrium fibers may have a cancellation effect on activation patterns. (a) Endocardium-epicardium fiber angle difference map. Red represents Aniso (anisotropic) regions, blue represents Iso (isotropic) regions. These Iso and Aniso regions are small and scattered on the atrium. An activation wave travel across the left atrium, for example, along the red dashed arrow, will pass through Iso and Aniso regions. (b) Conduction velocity increases and decreases along a long path on the atrium, however, these velocity variations did not accumulate.}
\label{fig:cancellation effect}
\end{figure}

Figure \ref{fig:cancellation effect}(a) show an endocardium-epicardium fiber angle difference map. Red represents 0 degree angle difference between the two layer's of fiber, and blue represents 90 degrees angle difference. Where the two layer's fiber angle difference is small (red region), activation wave traveling speed varies: it will be faster if wave direction is parallel to fiber direction, and slower if wave direction is perpendicular to fiber direction. We name these red regions Aniso (anisotropic) regions. On the contrary, in the blue regions, activation wave traveling speed does not vary much regardless of wave direction, we name these regions Iso (isotropic) regions.

For an activation wave that travels across the entire left atrium along a path as indicated by the red dashed arrow in Fig. \ref{fig:cancellation effect}(a), it crosses several Iso and Aniso regions. Activation wave traveling speed will not change much through the Iso regions, but it can increase or decrease through the Aniso regions. These speed increases and decreases mostly cancel each other, resulting in an overall near zero effect. Fig. \ref{fig:cancellation effect}(b) illustrates such speed changes along the path.

As a result of this cancellation effect, a left atrium with fiber organizations can be approximated by a left atrium that has isotropic conduction, with the conduction tuned to patient-specific value. This allows for an accurate fiber-independent model for the left atrium arrhythmia simulation.

\subsection{Stable vs meandering rotors}
\begin{figure*}[!ht]
\centering
\includegraphics[width = 1\textwidth]{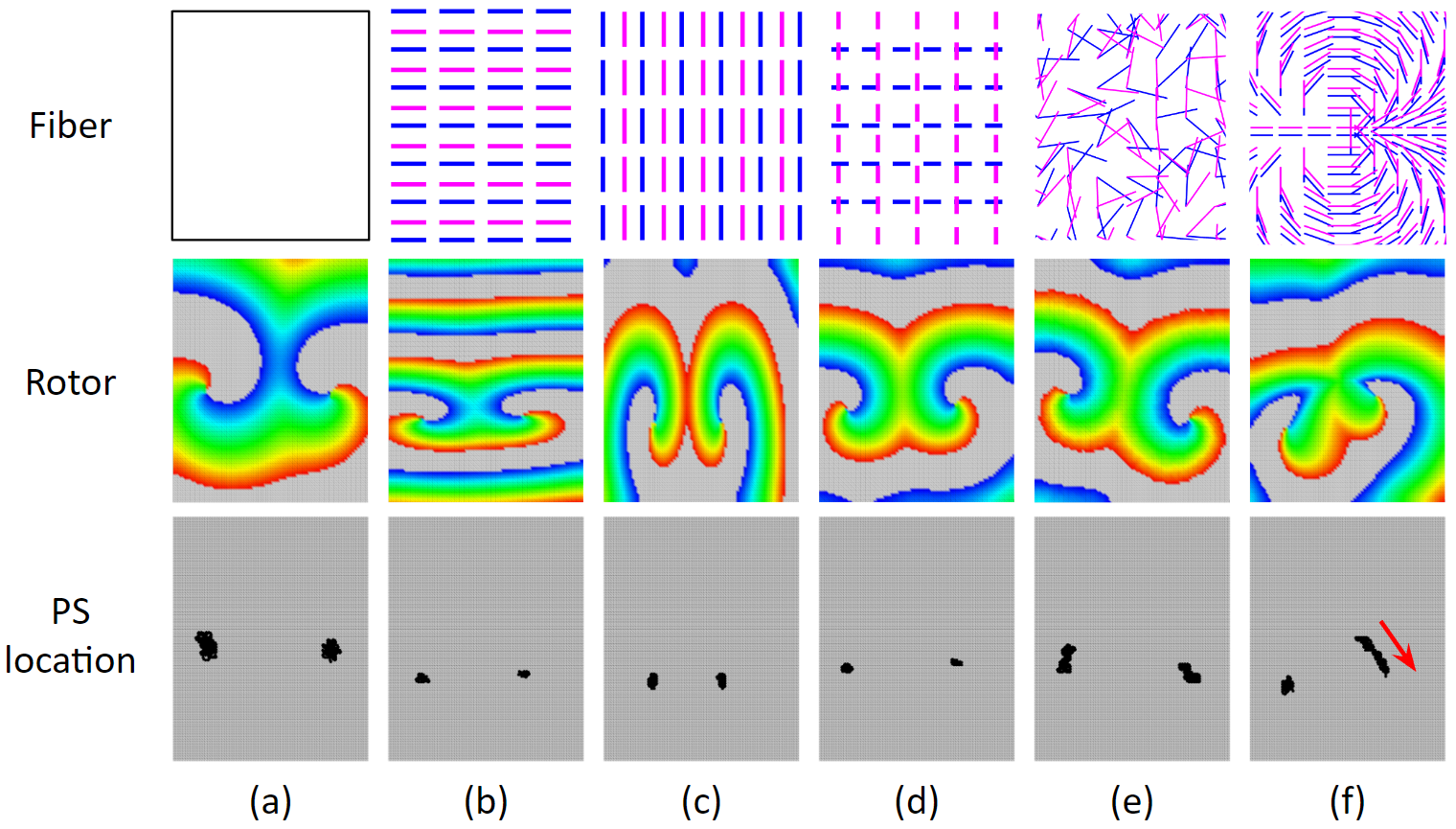}
\caption{Rotor arrhythmia experiments on a slab with different fiber organizations. Slab size is 80 mm $\times$ 100 mm $\times$ 2 mm. Each simulation produces 4 seconds of rotor arrhythmia. Row 1 are fibers: Blue represents endocardium fiber and magenta represents epicardium fiber. Column (a) has no fiber; column (b) fibers are in x direction; column (c) fibers are all in y direction; For column (d), endocardium fiber are in the x direction, epicardium fibers are in the y direction; column (e) has random fibers; and column (f) has curved fibers. Row 2 shows activation movie screenshots at time of 2600 ms. Row 3 shows the trajectories of the phase singularities (PS). Location tracking of the PS begins after the rotors finished their first rotation, ends 3 seconds or about 17 rotations later. We can see that PS locations moves the most in (f) because of the continuously changing fiber gradients.}
\label{fig:rotor slab experiment}
\end{figure*}

We conduct experiments on a 2D plane with different fiber orientations as depicted in Fig. \ref{fig:rotor slab experiment}. The \ac{PS} location (or rotor center location) plots indicate that rotors are relatively stable in cases (a)-(d), while they exhibit more meandering in cases (e) and (f). Method for detecting \ac{PS} locations is described in Appendix \ref{app:ps detection}. Studies have shown that fiber gradients can cause rotor to meander \cite{Rogers1994}, which explains why rotors move more in case (f) where the curved fibers have larger gradients.

\begin{figure}[!ht]
\centering
\includegraphics[width = 0.45\textwidth]{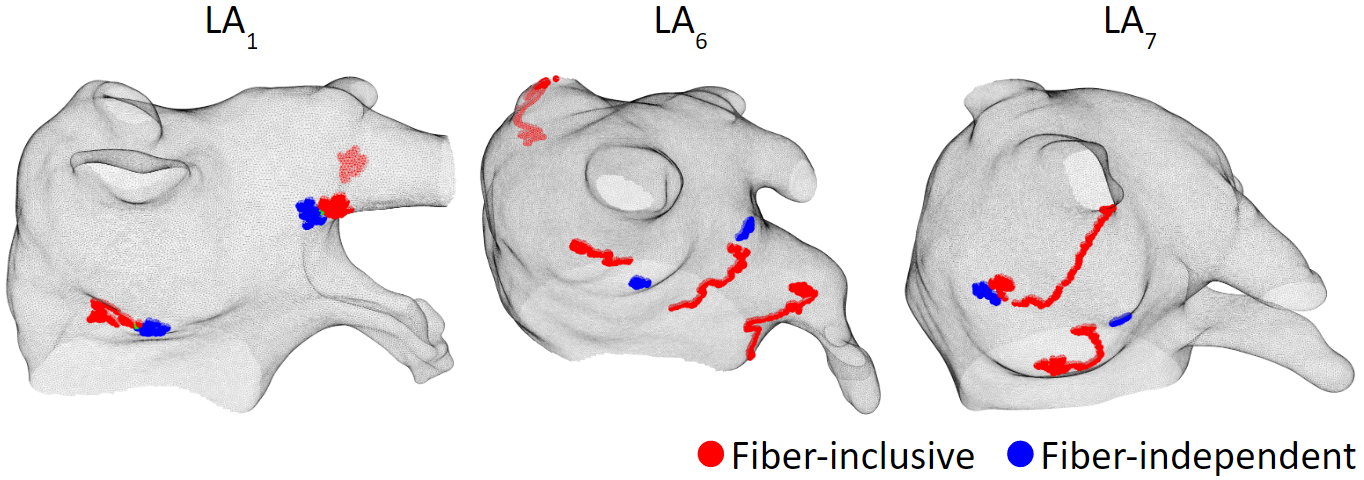}
\caption{Phase singularity locations for un-anchored rotors. The initial location of the rotors of the two models are similar. But as time progresses, rotors can meander away in fiber-inclusive models. However, un-anchored rotors in fiber-independent models are still stable.}
\label{fig:unanchored rotor}
\end{figure}

In the left atrium, the combination of fiber orientations and atrial geometry curvature may create large fiber gradients that can have a significant impact on rotor meandering. We run un-anchored rotor arrhythmias on all 7 left atria. The average distances of the nearest fiber-inclusive and fiber-independent rotor pairs range from 5 to 19 mm, and rotors often meanders further away as time progresses. Some of the \ac{PS} trajectories of a 3-second interval are shown in Fig. \ref{fig:unanchored rotor}. We can see that un-anchored rotors can meander long distances in fiber-inclusive models, but un-anchored rotors are still quite stable in fiber-independent models. This shows that fiber-independent models have a weak predictability on un-anchored rotors. However, clinically observed rotors rarely last more than a few rotations \cite{Fedorov2018}, and research found that stable rotors are ablation targets \cite{Narayan2013, Krummen2015, Kawata2013, Calvo2017}. 

\subsection{Clinical validation}

Our previous research \cite{He2021} has shown that it is possible to construct a model of the left atrium with a patient-specific geometry that reproduces clinical electroanatomical mapping data without incorporating fiber organization. We evaluated data from 15 patients and found that with appropriate fitting of the model, the average absolute \ac{LAT} error could be relatively small (approximately 5.5 ms for sinus rhythm and 11 ms for tachycardia). The average correlation of the activation maps generated by the fitted models and real electroanatomical maps was also relatively high (0.95 for sinus rhythm and 0.81 for tachycardia, respectively).

However, these findings did not answer the key question whether the resulting models have any predictive power. Answering this question would require a dedicated clinical study, which was not feasible at that time. Here, as an intermediate step, we carry out such a study in-silico. We hope that encouraging results presented here would provide a strong justification for the next step: A dedicated clinical study that would determine the utility of fiber-independent patient-specific models of the left atrium.

\subsection{Model tuning time}
The model tuning process takes approximately 18 seconds on a personal computer with an Intel Core i7-8700 CPU (3.20GHz) and an Nvidia GeForce GTX 1080 GPU. The required patient data (the electroanatomical map) is typically acquired at the beginning of an ablation procedure and usually takes 3 minutes to obtain with the Carto3 System. Therefore, the total time to tune our model can be as low as 3.3 minutes, which does not add significant burden in clinical practice, considering that an arrhythmia ablation procedure typically runs for 6 to 8 hours.

\section{Limitations}
We find that our fiber-independent left atrium model can accurately reproduce patient-specific focal and stable rotor arrhythmias. However, there are limitations to this finding. 

We do not consider scars in this study, which could play an important role in atrial fibrillation dynamics \cite{Gonzales2014}. However, since 79\% of first time ablation patients do not have scars \cite{Verma2005}, our model can still be helpful for clinical practice. 

In cases where scars are present, it is necessary to use individually tuned diffusion coefficients for each Cartesian node. This was demonstrated in our previous research, where we showed that a fiber-independent model can accurately reproduce activation patterns in left atria with scars \cite{He2021}.

The model is not equipped to predict the dynamics of meandering rotors. However, this limitation may not be as critical in clinical setting, given that only stable rotors are the primary ablation targets.

Among the several parameters, we only tune the diffusion coefficient. For most of the tachycardia, flutter, and macro re-entry cases, the action potential parameters we choose will work well. However, for more complex rhythms such as atrial fibrillation, due to its short cycle length, an accurate tuning of the action potential parameters may be crucial. 

The Mitchell-Schaeffer model we use is not a detailed ionic model, and it may not be a good choice to model complex rhythms such as atrial fibrillation. Still such a two-component model is good for modeling periodic arrhythmias. 

It is well established that more detailed bi-domain models are required for accurate simulation of electrical activity in the immediate vicinity of the stimulating electrodes and for modelling electrical defibrillation \cite{Roth2021}. However, we utilized the computationally more efficient mono-domain model. With regard to accuracy, the bi-domain models have no advantages over mono-domain models for simulating action potential propagations \cite{Potse2006}.

We simplify fiber organization into only two layers. The real left atrium has many more layers, and the number of layers also varies in different regions, as does the atrial thickness. If more layers were incorporated, we would need to study if the effects of fibers would become stronger.

We evaluate our model on the left atrium, because the most common atrial fibrillation sources are in the left atrium, therefore, it has more available clinical electroanatomical mapping data than the right atrium. 

\section{Conclusion}
We show that 1) fiber-independent left atrium model with tuned conduction can produce accurate activation maps of focal and stable rotor arrhythmias. 2) The model can be tuned to be patient-specific in a less than 4 minutes and can run in real-time to identify potential ablation targets during ablation procedure.

\appendices
\begin{appendices}
\section{Spatial and temporal resolutions}
\label{app:resolution}

\begin{figure}[!ht]
\centering
\includegraphics[width = 0.45\textwidth]{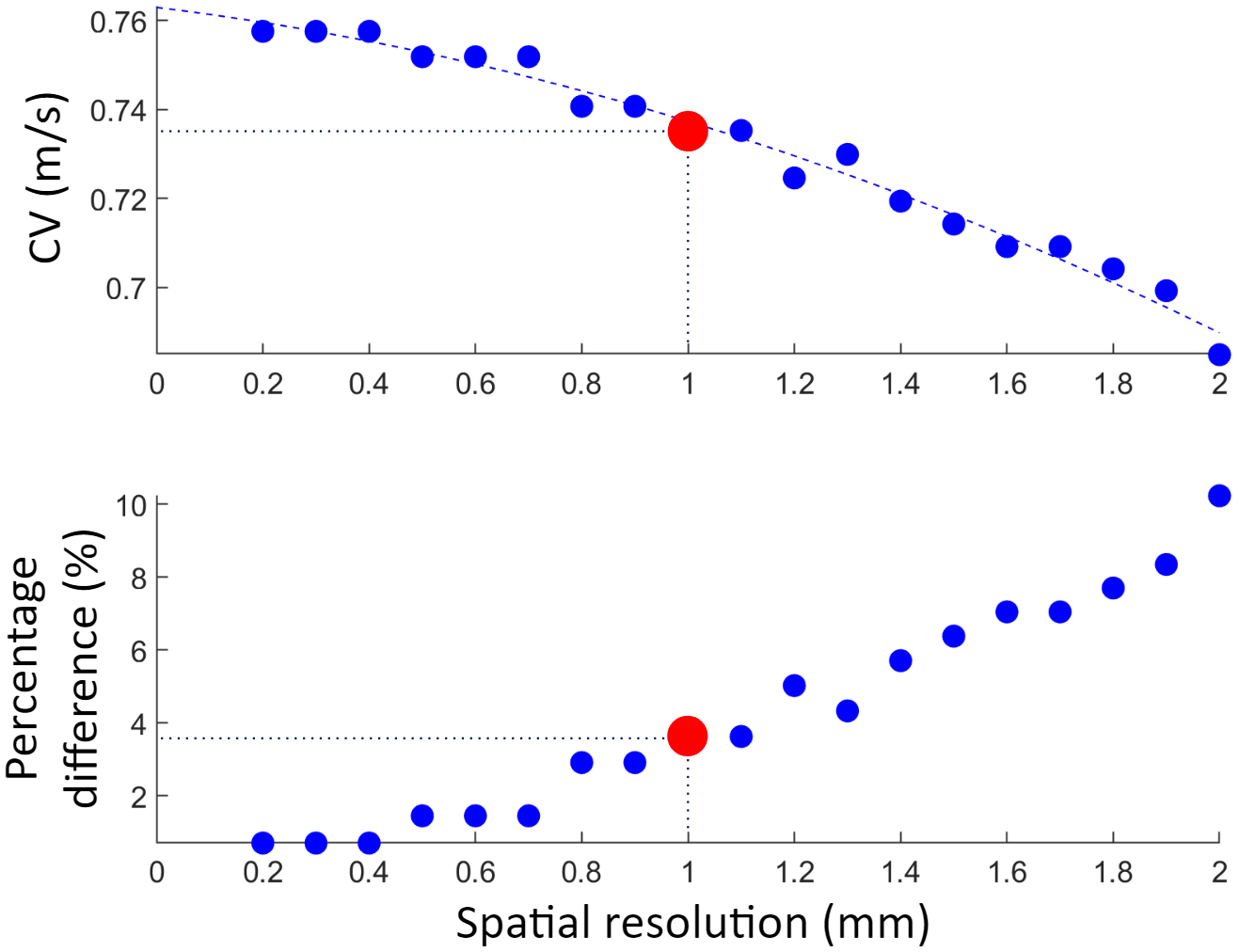}
\caption{Top: Conduction velocity (CV) as a function of the discretization step. The dashed blue line is a quadratic fit. Bottom: Relative deviation of \ac{CV} at a given discrtization step from the asymptotic value of \ac{CV} at 0 mm discretization step at different resolutions. For spatial resolution of 1 mm and temporal resolution of 0.01 ms (the big red dot), the \ac{CV} percentage difference is 3.6\%.}
\label{fig:delta vs conduction velocity}
\end{figure}

We run simulations on a slab of 100 mm $\times$ 4 mm $\times$ 4 mm. (The long length of the slab is to help increase \ac{CV} computational accuracy.) The heart model parameters are chosen such that the \ac{CV} values are close to the physical values \cite{Harrild2000}. Assume isotropic conduction, or no fiber organization. Spatial resolutions are set to 0.2, 0.3, ..., 2.0 mm. Temporal resolution $dt$ is set to 0.01 ms. Results are summarized in Fig. \ref{fig:delta vs conduction velocity}. In this paper, spatial resolution is 1 mm and temporal resolution is 0.01 ms. Using such resolutions, the accuracy of \ac{CV} is adequate with a deviation from the asymptotic value of 3.6\%, which is much smaller than the usual required 10\%.

\section{Solving the heart model differential equations}
\label{app:solve heart model}
To solve the differential equations \eqref{eq:mitchell schaeffer}, initial values at t = 0 are given ($u_0 = 0$ and $h_0 = 1$), then solutions of the next time step are computed using the explicit Euler method as shown in \eqref{eq:explicit Euler}. The time step is $\Delta t = 0.01$ ms (justification is in Appendix \ref{app:resolution}). 

\begin{align}
\label{eq:explicit Euler}
\begin{split}
u_{t+1} &= \left( \frac{h_{t}(u_{t})^2(1-u_{t})}{\tau _{in}}-\frac{u_{t}}{\tau _{out}}+J+\bigtriangledown \cdot (D\bigtriangledown u_{t}) \right) \Delta t + u_{t}
\\
h_{t+1} &= \left\{\begin{matrix}\ \frac{1-h_{t}}{\tau_{open}} \Delta t + h_{t} \ \text{if}\ u_{t}<u_{gate} \\ \ \frac{-h_{t}}{\tau_{close}} \Delta t + h_{t} \ \text{if}\ u_{t} \geq u_{gate}\end{matrix}\right.
\end{split}
\end{align}

\begin{figure}[!ht]
\centering
\includegraphics[width = 0.3\textwidth]{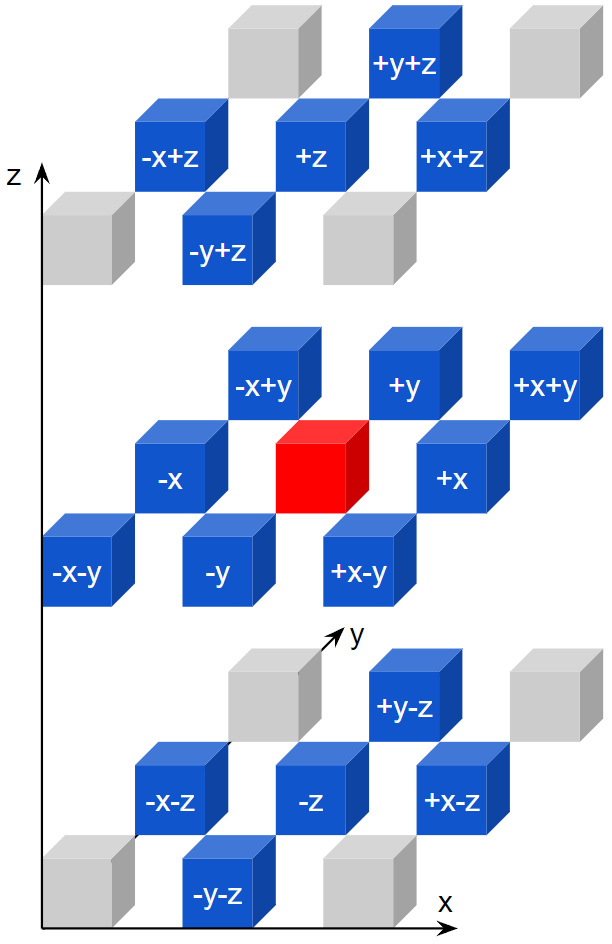}
\caption{Node neighbors. The red node has 26 neighbors, but only 18 of them are involved in solving the heart model equations, and they are colored blue. If the red node has coordinate $(x,y,z)$, then the neighbor blue node labelled "-x+y" has coordinate $(x-\Delta,y+\Delta,z)$.}
\label{fig:node neighbors}
\end{figure}

To compute the diffusion term $\bigtriangledown \cdot (D\bigtriangledown u)$, we follow \cite{McFarlane2010} that uses a 19-node stencil as shown in Figure \ref{fig:node neighbors}.

Give $D$ a matrix notation as \eqref{eq:D2}:

\begin{align}
\label{eq:D2}
D = d \left ( rI+\left ( 1-r \right )ff^\top \right ) = d \begin{bmatrix} D_{11} & D_{12} & D_{13} \\
D_{21} & D_{22} & D_{23} \\
D_{31} & D_{32} & D_{33} \end{bmatrix}
\end{align}

Then the diffusion term becomes:

\begin{align}
\label{eq:D3}
\bigtriangledown \cdot \left ( D\bigtriangledown u \right ) = d \left( \sum_{i=1}^{3}\sum_{j=1}^{3}\left ( \frac{\partial D_{ij}}{\partial x_i} \frac{\partial u}{\partial x_j} + D_{ij}\frac{\partial^2 u}{\partial x_i \partial x_j} \right ) \right)
\end{align}

Where $x_1 = x$, $x_2 = y$ and $x_3 = z$. In the following equations, we introduce a subscript notation: $\square_{...}$ represents the value for the node at coordinate $(x,y,z)$, which is the red node in Figure \ref{fig:node neighbors}; $\square_{.+-}$ represents the value for the node at coordinate $(x,y+\Delta,z-\Delta)$, which is the blue node labelled "+y-z" in Figure \ref{fig:node neighbors}. The partial derivatives can be approximated as \eqref{eq:derivatives}, and other terms in \eqref{eq:D3} can be approximated in similar manners. 

\begin{align}
\label{eq:derivatives}
\begin{split}
\frac{\partial u_{...}}{\partial x} &\approx \frac{u_{+..} - u_{-..}}{2 \Delta} \\
\frac{\partial^2 u_{...}}{\partial x^2} &\approx \frac{u_{+..}-2u_{...}+u_{-..}}{\Delta^2} \\
\frac{\partial^2 u_{...}}{\partial x \partial y} &\approx \frac{ u_{++.} - u_{-+.} - u_{+-.} + u_{--.} } {4\Delta^2}
\end{split}
\end{align}

\subsection{Boundary conditions}
As shown in Figure \ref{fig:node neighbors}, to compute the diffusion term for a node, it involves 18 neighbor nodes. However, some nodes may not have all those 18 neighbors because it is at the boundary. For those none-existing neighbors, we can multiply 0 to eliminate the associated terms, to satisfy the no-flux (Neumann) boundary conditions. Substitute \eqref{eq:derivatives} into \eqref{eq:D3}, rearrange the terms, and add in an indicating variable $\delta$, we then have an expression of the diffusion term with implicit no-flux boundary conditions. The indicator variable $\delta = 1$ if there is a Cartesian node at that coordinate specified in the subscript, and $\delta = 0$ if it is void at that coordinate. 

\subsection{The indicator variable $\delta$}
Each node has an index $i$, typically, for a left atrium, the total number of nodes is about $N=100,000$. To find out $\delta$, we first find out the neighboring nodes indices. Create a $N \times 18$ matrix $neighbor\_id$, in which each row $i$ stores node $i$'s 18 neighbors (as shown in Figure \ref{fig:node neighbors}) in sequence: these neighbors are "+x", "-x", ..., "-x-z" as shown in Table \ref{tb:node neighbors}. For example, if we want to find the "-x+z" neighbor of a node at $(x,y,z)$, than check which node has coordinate $(x-\Delta,y,z+\Delta)$. 

For programming in Matlab, variable indices starts at 1, therefore $i$ starts at 1 and ends at N. If there is a node at that coordinate, then record that node's index; if there is not a node, then record 0. For example, the 120th row of $neighbor\_id$ may be as Table \ref{tb:node neighbors}. Take the sign of $neighbor\_id$ to get the value of $\delta$:

\begin{align}
\label{eq:delta}
\delta = sign(neighbor\_id)
\end{align}

\begin{table}[!ht]
    \centering
    \caption{An example of a row of $neighbor\_id$}
    \begin{tabular}{ cccccc }
    \hline
        $+x$ & $-x$ & $+y$ & $-y$ & $+z$ & $-z$ \\
        201 & 39 & 121 & 119 & 0 & 0 \\\hline\hline
        $+x+y$ & $-x+y$ & $+x-y$ & $-x-y$ & & \\
        202 & 40 & 200 & 38 & & \\\hline\hline
        $+y+z$ & $-y+z$ & $+y-z$ & $-y-z$ & & \\
        0 & 0 & 0 & 0 & & \\\hline\hline
        $+x+z$ & $-x+z$ & $+x-z$ & $-x-z$ & & \\
        0 & 0 & 0 & 0 & & \\
    \hline
    \end{tabular}
    \label{tb:node neighbors}
    \begin{flushleft}
    The neighbor nodes indices of node $i=120$ may look like this table. It means, for example, its "-x" neighbor is node $i=39$, and there is no neighbor at "-y+z". 
    \end{flushleft}
\end{table}

\subsection{Programming implementation of the diffusion term}
The computer programming implementation of the diffusion term $\bigtriangledown \cdot ( D\bigtriangledown u )$ is \eqref{eq:diffusion term with boundary condition simplified}.

\begin{align}
\label{eq:diffusion term with boundary condition simplified}
\begin{split}
&\frac{d}{4\Delta^2} \{ P_1 (u_{+..}-u_{...}) + P_2 (u_{-..}-u_{...}) + P_3 (u_{.+.}-u_{...}) + \\
&P_4 (u_{.-.}-u_{...}) + P_5 (u_{..+}-u_{...}) + P_6 (u_{..-}-u_{...}) + \\
&P_7 (u_{+..}-u_{-..}) + P_8 (u_{.+.}-u_{.-.}) + P_9 (u_{..+}-u_{..-}) + \\ 
&P_{10} (u_{++.}-u_{+-.}) + P_{11} (u_{--.}-u_{-+.}) + \\
&P_{12} (u_{+.+}-u_{+.-}) + P_{13} (u_{-.-}-u_{-.+}) + \\
& P_{14} (u_{.++}-u_{.+-}) + P_{15} (u_{.--}-u_{.-+}) \}
\end{split}
\end{align}

Where

\begin{align}
\label{eq:parts}
\begin{split}
P_1 = &4\delta_{+..} D_{...}^{11}, P_2 = 4\delta_{-..} D_{...}^{11}, P_3 = 4\delta_{.+.} D_{...}^{22} \\
P_4 = &4\delta_{.-.} D_{...}^{22}, P_5 = 4\delta_{..+} D_{...}^{33}, P_6 = 4\delta_{..-} D_{...}^{33} \\
P_7 = &\delta_{+..}\delta_{-..} [\delta_{+..}\delta_{-..}(D_{+..}^{11}-D_{-..}^{11}) +\\
&\delta_{.+.}\delta_{.-.}(D_{.+.}^{21}-D_{.-.}^{21}) + \delta_{..+}\delta_{..-}(D_{..+}^{31}-D_{..-}^{31})] \\
P_8 = &\delta_{.+.}\delta_{.-.} [\delta_{+..}\delta_{-..}(D_{+..}^{12}-D_{-..}^{12}) +\\
&\delta_{.+.}\delta_{.-.}(D_{.+.}^{22}-D_{.-.}^{22}) + \delta_{..+}\delta_{..-}(D_{..+}^{32}-D_{..-}^{32})] \\
P_9 = &\delta_{..+}\delta_{..-} [\delta_{+..}\delta_{-..}(D_{+..}^{13}-D_{-..}^{13}) + \\
&\delta_{.+.}\delta_{.-.}(D_{.+.}^{23}-D_{.-.}^{23}) + \delta_{..+}\delta_{..-}(D_{..+}^{33}-D_{..-}^{33})] \\
P_{10} = &2\delta_{++.}\delta_{+-.} D_{...}^{12}, P_{11} = 2\delta_{--.}\delta_{-+.} D_{...}^{12} \\
P_{12} = &2\delta_{+.+}\delta_{+.-} D_{...}^{13}, P_{13} = 2\delta_{-.-}\delta_{-.+} D_{...}^{13}\\
P_{14} = &2\delta_{.++}\delta_{.+-} D_{...}^{23}, P_{15} = 2\delta_{.--}\delta_{.-+} D_{...}^{23}
\end{split}
\end{align}

\section{Phase singularity detection}
\label{app:ps detection}
Phase singularity (PS) refers to a pivot point around which an activation wave rotates. To identify PS, the phase values of the three vertices of a triangle on the left atrium mesh are analyzed. If specific criteria are met, the triangle is determined to contain a PS.

The phase values of a vertex is a time sequence that linearly increases from 0 to 2$\pi$. It starts at the beginning of an activation and ends at the beginning of the subsequent activation.

\begin{figure}[!ht]
\centering
\includegraphics[width = 0.3\textwidth]{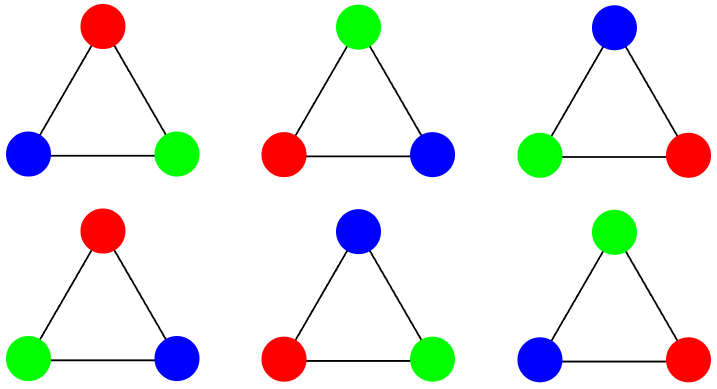}
\caption{The six scenarios that a triangle contains a phase singularity.}
\label{fig:ps}
\end{figure}

For each time instance, we transform the phase value $p$ into one of three colors:
\begin{itemize} 
\setlength\itemsep{0.5em}
  \item If ($p \geq 0$ and $p < \frac{1}{6}2\pi$) or ($p > \frac{5}{6}2\pi$ and $p \leq 2\pi$), assign red, or $[1\ 0\ 0]$ in RGB representation.
  \item If $p \geq \frac{1}{6} 2\pi$ and $p<\frac{1}{2} 2\pi$, assign green, or $[0\ 1\ 0]$.
  \item If $p \geq \frac{1}{2} 2\pi$ and $p \leq \frac{5}{6} 2\pi$, assign blue, or $[0\ 0\ 1]$.
\end{itemize}
For a triangle, we can create a $3\times3$ matrix that each row is the RGB representation of a vertex's color. This matrix can then be transformed so that different combinations of vertex colors are converted into different numerical values:
\begin{equation}
\begin{split}
\label{eq:matrix to number}
&[3^2 \ 3^1 \ 3^0] \begin{bmatrix} c_{11} & c_{12} & c_{13} \\ c_{21} & c_{22} & c_{23} \\ c_{31} & c_{32} & c_{33} \end{bmatrix} \begin{bmatrix} 0 \\ 1 \\ 2 \end{bmatrix} \\
=& 9c_{12} + 18c_{13} + 3c_{22} + 6c_{23} + c_{32} + 2c_{33} 
\end{split}
\end{equation}
If a triangle contains a PS, then the numerical value will be one of these: 7, 11, 21, 5, 19, and 15. And these six scenarios are depicted in Fig. \ref{fig:ps}. 

Performing PS detection for every time instance on all triangles of the mesh will result in PS trajectories, representing the movement of the PS over time.

Note that it is a property of the triangular mesh that the sequence of the three vertices of a triangle follows the right-hand-rule: right hand four fingers curl following the vertices sequence, then the thumb will represent the face normal that points outwards of the atrium mesh. This ensures the detected PS rotational direction (clockwise or counterclockwise) is correct.

\end{appendices}

\bibliographystyle{ieeetr}
\bibliography{reference}

\end{document}